\documentclass{article}
\usepackage[a4paper, total={6in, 8in}]{geometry}
\usepackage{calrsfs}
\usepackage{hyperref}
\usepackage{color}
\usepackage{float}
\usepackage{esvect}
\usepackage{booktabs}
\usepackage{siunitx}
\usepackage{authblk}
\usepackage{calrsfs}
\usepackage{amsmath}
\usepackage{amssymb}
\usepackage{braket}
\usepackage[version=4]{mhchem}
\usepackage{mathtools}          

\usepackage{graphicx}
\usepackage{dcolumn}
\usepackage{graphicx}
\usepackage{bm}
\usepackage{sectsty}
\usepackage{graphicx}

\sectionfont{\fontsize{12}{15}\selectfont}

\begin{document}

\title{Determination of the coefficient of thermal expansion by measuring frequency of a heated music wire}
\author[1]{\small Robert D. Polak}
\author[1]{\small Michael R. Harris}
\author[1]{\small Kiet A. Nguyen}
\author[1]{\small Anthony Kearns}
\affil[1]{\small Department of Physics, Loyola University Chicago, Chicago, IL 60660, USA}




%


%
%
%


\date{}

\maketitle

\section*{}

 \hspace{5mm}Engaging with physical and material properties through empirical observation is a fundamental part of undergraduate physics and engineering education. Several works have proposed experiments to determine thermal physical constants of materials such as finding the coefficient of linear expansion [1-4]. As Dajbych [5] and Polak et al. [6] have shown, methods for experimentally verifying physical constants can be done by measuring the frequency of a plucked high-carbon steel wire on a guitar. Building upon our previous work [6,7], we have extended our method to verify the coefficient of linear thermal expansion, $\alpha_T$, through an accessible procedure directed at introductory physics education. To do this, we heated a guitar wire by running a current through it, causing the string to expand and resulting in a measurable decrease in frequency. Using only a DC power source, an acoustic guitar, 2 digital multimeters, and a cellphone, students can calculate the coefficient of thermal expansion within a typical lab duration.

The frequency of a stringed instrument will change with increasing temperature from reduced tension on the string as a result of thermal expansion. The frequency $f$ is related to tension through
\begin{equation}
    f\lambda = v = \sqrt{\frac{F_{\perp}}{\mu}},
\end{equation}
where $\lambda$ is the wavelength of the wave, $v$ is the speed of the wave, $F_{\perp}$ is the tension on the string and $\mu$ is the mass per unit length of the string. We can ignore the temperature dependence of  $\mu$, as it changes less than 0.1 $\%$ over the temperature range used in this work. The tension $F_{\perp}$ can then be written as a function of the temperature $T$
\begin{equation}
    F_{\perp}(T) = F_{\perp}(T=T_o) - \Delta F_{\perp}(\Delta T) ,  
\end{equation}
where $T_o$ is the initial temperature and $ \Delta F_{\perp}$ is the change in tension due to a change in temperature $\Delta T = T - T_o$. The change in temperature can be calculated from the change in resistance in the wire from 
\begin{equation}
\Delta T = \frac{\frac{R}{R_o} - 1}{\alpha_R},
\end{equation}
where $R$ is the resistance at temperature, $T$, $R_o$ is the resistance at temperature, $T_o$, and $\alpha_R$ is the temperature coefficient of resistivity for the wire material.
The change in tension is due to thermal expansion and is related to the linear expansion of the string, $\Delta L(\Delta T)$, by
\begin{equation}
    \Delta F_{\perp}(\Delta T) = \frac{YA \Delta L(\Delta T)}{L_o},
\end{equation}
where $Y$ is the Young's modulus of the string, $A$ is the cross-sectional area and $L_o$ is the length of the string at temperature $T_o$. The change in length is given by
\begin{equation}
    \frac{\Delta L(\Delta T)}{L_o}=\alpha_T \Delta T ,
\end{equation}
where $\alpha_T$ is the coefficient of thermal expansion of the material. Combining Eqs. (1), (2), and (4) while noting that $\lambda=2L_o$ and the density of the string $ \rho=\frac{\mu}{A} $, we find the fundamental frequency to be
\begin{equation}
    f^2 = f_o^2 - \frac{Y\alpha_T}{4L_o^2 \rho} \Delta T ,
\end{equation}
where $f_o$ is the frequency measured at $T_o$. By measuring the frequency $f$ of a plucked guitar string as a function of temperature, we can use the slope of the linear fit of $f^2$ vs. $\Delta T$ to find $\alpha_T$. 
To perform the experiment, we replaced the outermost string of an acoustic guitar with a high-carbon steel music wire. The music wire was heated by passing an electric current through it. We created the circuit shown in Fig. 1 by connecting the music wire to the power supply below of the guitar's saddle and above the guitar's nut, as shown in Fig. 2. The ammeter and voltmeter measured the current through and voltage across the wire respectively, while frequency was measured using the G-Strings mobile application [8] on a cellular phone held about $3$ inches from the wire, as shown in Fig. 3.

We first plucked the string with no current applied and measured its frequency. This measurement is not used in calculations, but serves as a check that the guitar string has enough tension on it and that the frequency of the heated string is lower. In our experiment, the frequency at room temperature was about 220 Hz when using a 0.016 inch diameter wire. We then applied a current of $0.5A$ to the string and waited 30 seconds for it to reach thermal equilibrium [9]. We recorded the current and voltage through the wire, then plucked the string once and recorded its frequency. This frequency is $f_o$ measured at the temperature, $T_o$. We used these voltage and current values to find the resistance, $R_o$, at temperature $T_o$. We then increased the currents in steps of $0.1A$ up to about $1.6A$. Again, at each step, we waited 30 seconds for the wire to reach thermal equilibrium, recorded the voltage and current through the wire, then plucked the string once and measured its frequency. We found the change in temperature at each current by measuring the resistance of the wire and applying Eq. (3), using $\alpha_R= .0039 K^{-1}$ [7]. 

Figure 4 shows $f^2$ graphed as a function of $\Delta T$ for our experiment. By applying a linear fit, the slope will be $\frac{Y \alpha_T}{4L_o^2 \rho}$ and the y-intercept will be $f_o^2$ according to Eq. (5). This slope was found to be $m = \SI{-214.67}{\kelvin^{-1} \s^{-2}}$. Using $L_o = \SI{0.62}{\meter}$, $Y = \SI{1.98e 11}{\pascal}$ [6,10], and $\rho = \SI{7640}{\kg \meter^{-3}}$ [10] we calculated an $\alpha_T$ of $\SI{1.27e-5}{\per \kelvin}$, which is within $5 \%$ of the accepted $\alpha_T$ of $\SI{1.22e-5}{\per \kelvin}$ [11].

In designing this experiment, we found several issues that degraded the quality of our results. Foremost, we found that when $\Delta T$ grew larger than 50 K, Eq. (6) no longer gave a linear result. To address this issue, we chose the range and step size of applied currents to provide about 12 data points with $\Delta T$ not exceeding 40 $K$. Additionally, while analog power supplies could be used, the difficulty in creating precise changes in current proved difficult for producing a well formed dataset. Therefore, we used a BK Precision 1787B power supply for its digital current adjustment, leading to precise current increases. Further, because repetitive plucking causes the guitar string to cool, resulting in a higher frequency measurement, we determined it is crucial to measure the frequency on the first pluck of the string. If the first pluck fails to give a frequency measurement, it is necessary to wait 30 seconds to allow the wire to reach equilibrium again before another attempt is made.




When considered with our previous work regarding the verification of Young's modulus and the temperature coefficient of resistivity [6,7], our procedure provides a holistic approach to an empirical understanding of mechanical and physical material properties. We have shown that with standard equipment found in most university labs and inexpensive stringed instruments, students can develop a comprehensive understanding of fundamental thermal physics properties through an intuitive and hands-on procedure.

\section*{References}
1.  H. Fakhruddin, Phys Teach \textbf{44}, 82 (2006).\\
2.  S.S.R. Inbanathan, K. Moorthy and G. Balasubramanian, Phys Teach \textbf{45}, 566 (2007).\\
3.  R. Scholl and B.W. Liby, Phys Teach \textbf{47}, 306 (2009).\\
4.  E. H. Graf, Phys Teach \textbf{50}, 181 (2012).\\
5.  O. Dajbych, "Coefficient of Thermal Expansion of Electric Guitar String Material Determination" in Modern Methods of Constuction Design: Proceedings of ICMD 2013, 329-334 (2014).
\\
6.  R. D. Polak, A. Davenport, A. Fischer, and J. Rafferty, Phys Teach \textbf{56}, 122-123 (2018).\\
7. R. D. Polak, M. R. Harris, K. A. Nguyen, and A. Kearns. Manuscript Submitted for Publication. \\ 
8. Tuner - gStrings Free version 2.3.4, Available from
\url{https://play.google.com/store/apps/details?id=org.cohortor.gstrings&hl=en$_$US&gl=US}, Accessed: 2020\\
9. At $0.5 A$, the voltage applied to the wire is $0.53V$ and the resistance is $1.06\Omega$. At $1.6 A$, the voltage applied to the wire is $1.97 V$ and the resistance is $1.23\Omega$ . \\
10. Precision Brand Music Wire (0.016 inch diameter); UPC. No. 21016\\
11. Valves Instruments Plus Ltd., Available from \url{https://www.vip-ltd.co.uk/Expansion/Thermal_Expansion.pdf}, Accessed: 2022.\\

\pagebreak

\begin{figure}[h]
\centering
\includegraphics[scale = .4]{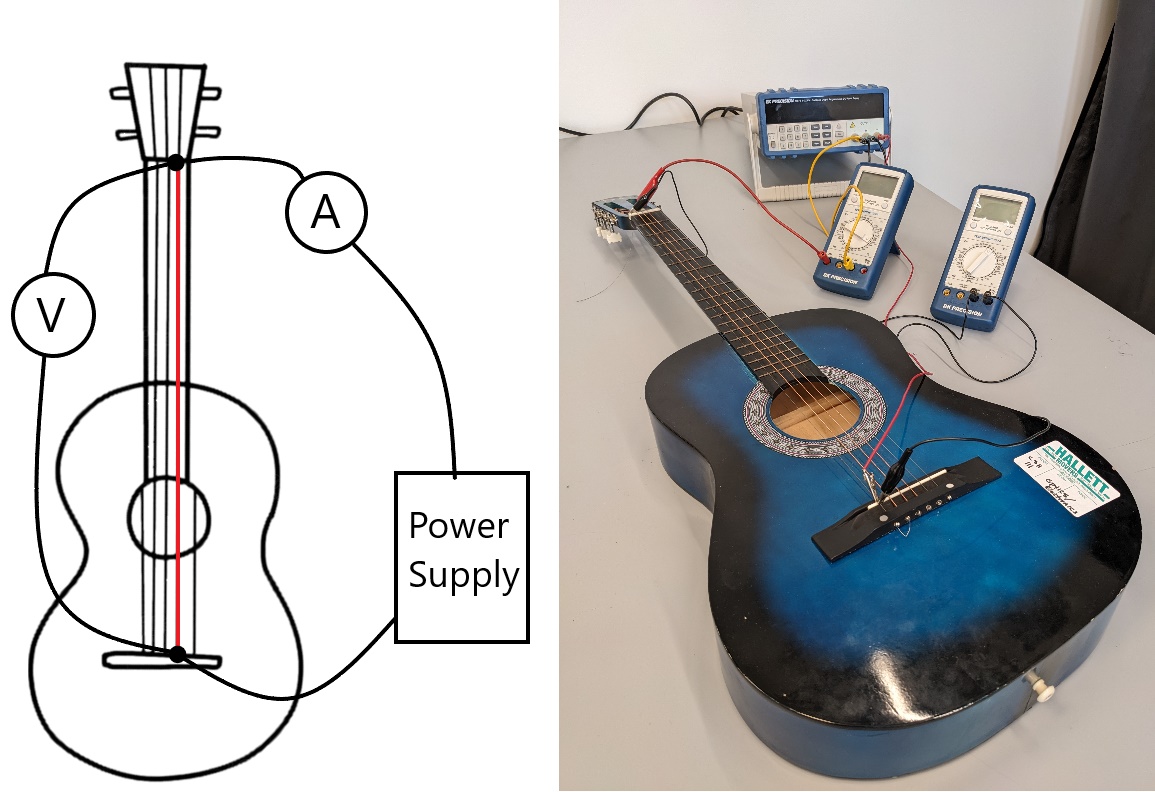}
\caption{A closed circuit created on a guitar with music wire, a power supply, and two multimeters to measure voltage and current.}
\end{figure}

\begin{figure}[h]
\centering
\includegraphics[scale = .5]{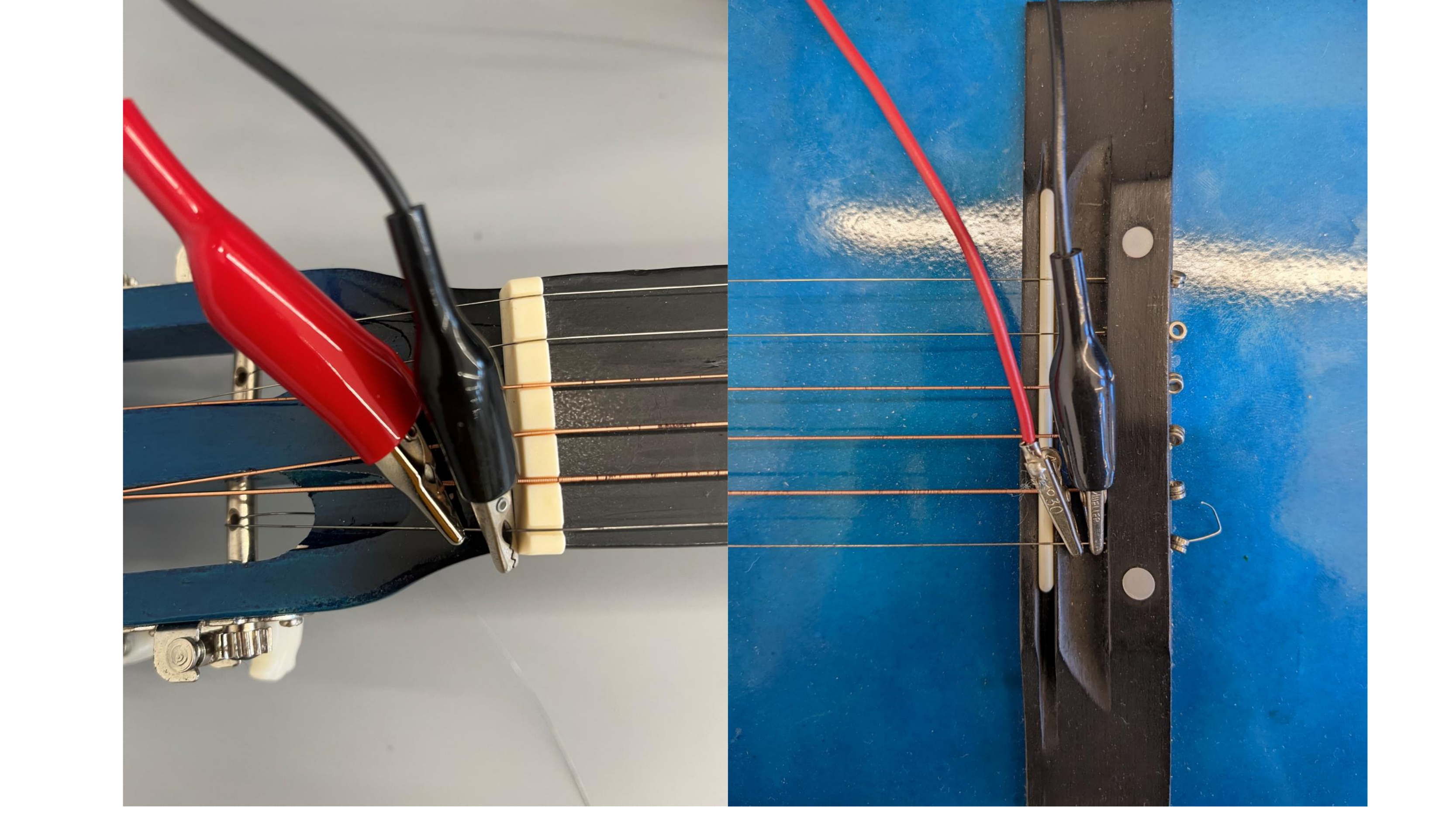}
\caption{The connection points on both ends of the guitar. All connections are placed above the nut (left) and below the saddle (right) to ensure that the electrical connections do not interfere with the harmonics of the plucked wire.}
\end{figure}

\begin{figure}[h]
\centering
\includegraphics[scale = .2]{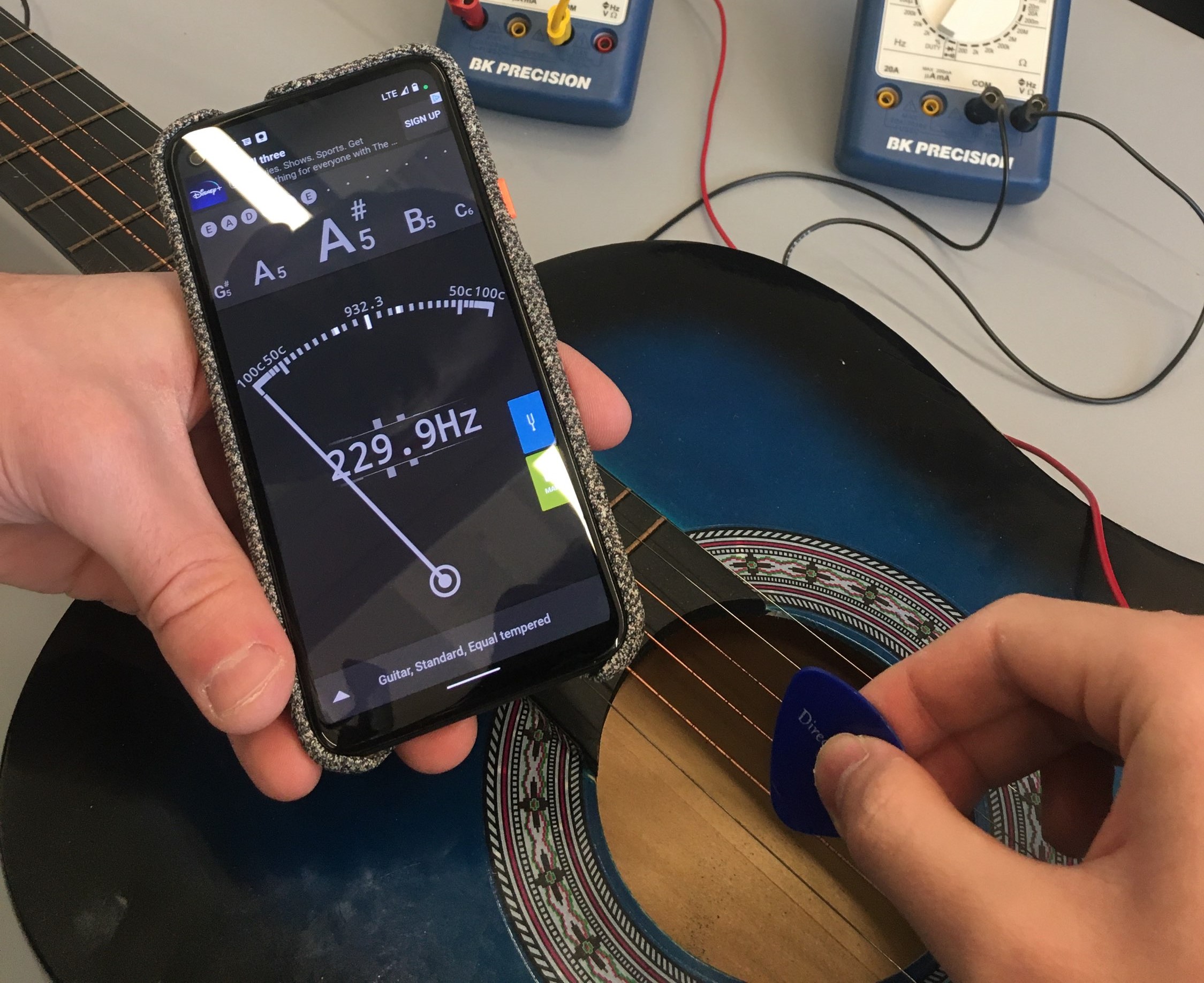}
\caption{The positioning of the mobile phone while collecting frequency measurements. The phone's microphone is approximately 3 inches from the plucked music wire.}
\end{figure}

\begin{figure}[h]
\centering
\includegraphics[]{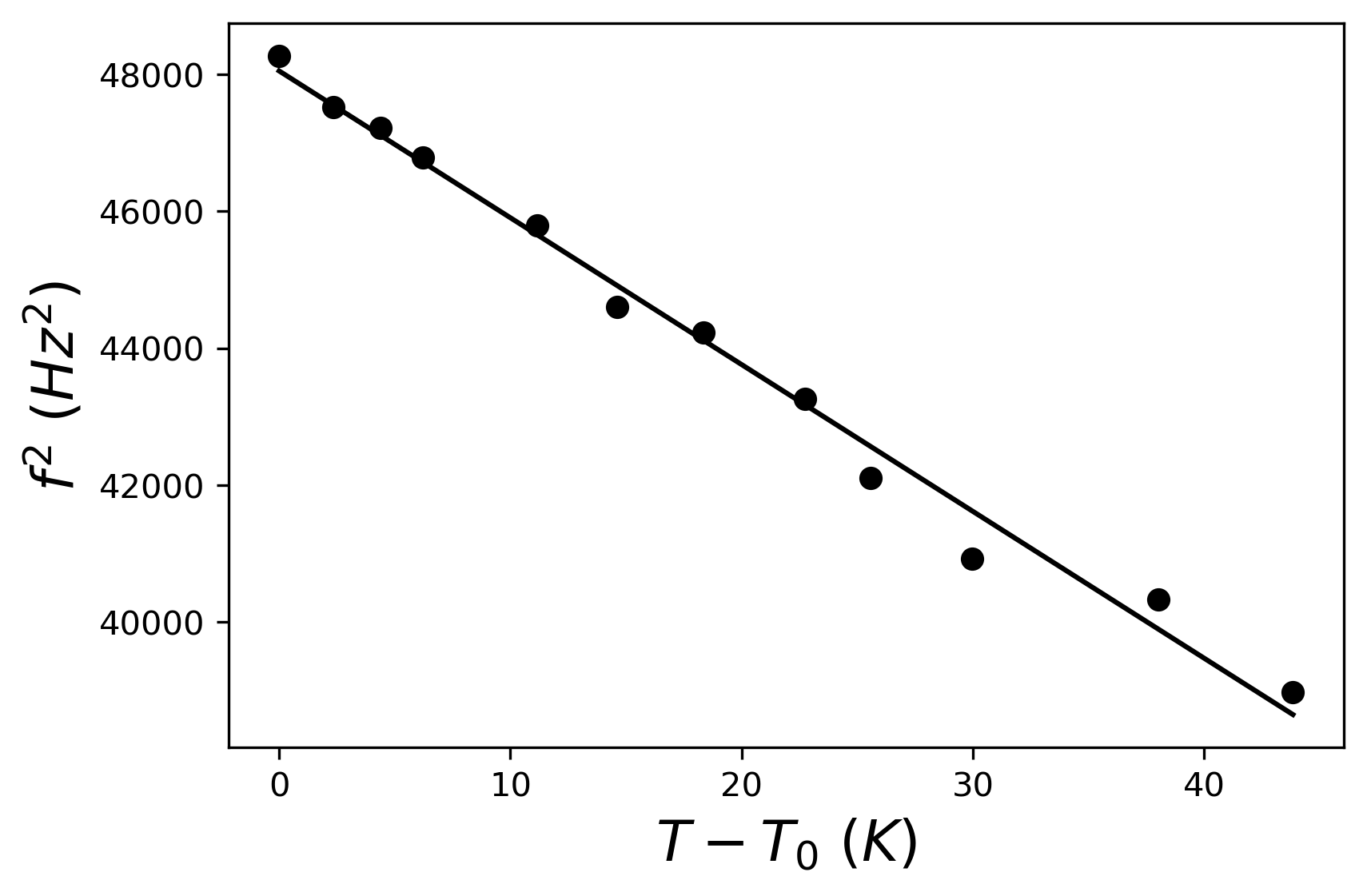}
\caption{Plot of frequency squared against the change in temperature for currents  from $0.5A$ to $1.6 A$. A linear fit of the data yields $\alpha_T = \SI{1.27e-5}{\per \kelvin}$.}
\end{figure}


\end{document}